# BLOCH OSCILLATIONS IN A JOSEPHSON CIRCUIT

N. BOULANT, G. ITHIER, F. NGUYEN, P. BERTET, H. POTHIER, D. VION, C. URBINA, AND D. ESTEVE

*Quantronics, Service de Physique de l'Etat Condensé, URA 2464*
*CEA-Saclay, 91191 Gif-sur-Yvette, France*

Bloch oscillations predicted to occur in current-biased single Josephson junctions have eluded direct observation up to now. Here, we demonstrate similar Bloch oscillations in a slightly richer Josephson circuit, the quantronium. The quantronium is a Bloch transistor with two small junctions in series, defining an island, in parallel with a larger junction. In the ground state, the microwave impedance of the device is modulated $2e$ periodically with the charge on the gate capacitor coupled to the transistor island. When a current $I$ flows across this capacitor, the impedance modulation occurs at the Bloch frequency $f = I/(2e)$, which yields Bloch sidebands in the spectrum of a reflected continuous microwave signal. We have measured this spectrum, and compared it to predictions based on a simple model for the circuit. We discuss the interest of this experiment for metrology and for mesoscopic physics.

## 1  Bloch oscillations in Josephson junctions

The phenomenon of Bloch oscillations [1], was first considered for a quantum particle moving in a periodic potential and subject to a constant driving force (see [2] for a review). When the particle stays in the first Bloch band, its quasi-momentum changes linearly with time until it reaches the boundary of the first Brillouin zone, where it is Bragg-reflected to the symmetric opposite band-edge and increases again. The velocity of the particle oscillates during the motion. This phenomenon pertains to many physical situations where a quantum system with a periodic potential is subject to a constant drive [2]. Experimental evidence of Bloch oscillations can be found in particular for electrons in solid state superlattices [3], and for ultracold atoms in optical lattices [4]. Bloch oscillations are also predicted to occur in a single Josephson junction biased by a dc current $I$ [5]. The variables describing the system are the charge Q on the junction capacitance and the phase difference $\delta$ across the junction. They form a set of conjugated variables: $[Q,\delta] = i(2e)$, with $e$ the electron charge. The Hamiltonian writes:

$$H = \hat{Q}^2/(2C) - E_J \cos(\hat{\delta}) - I\hat{\delta}/(2e), \qquad (1.1)$$

with C the junction capacitance, and $E_J$ the Josephson energy. The Bloch oscillations should manifest here as a periodic oscillation of the voltage across the junction at the Bloch frequency $f = I/(2e)$. The locking of these oscillations by an external microwave signal applied to the junction is predicted to induce voltage steps at constant current in the current-voltage characteristic. These steps would be dual of Shapiro steps in voltage biased Josephson junctions, and provide an appealing solution for the metrology of currents. This experiment is however difficult because a source impedance large compared to the resistance quantum $R_Q = h/(2e)^2 \simeq 6.5\,k\Omega$ is required over a wide frequency range. Although a Bloch-nose feature related to Bloch oscillations and an effect of microwaves have been observed [6-7], the locking regime has never been reached. In this work, we have performed a related experiment which shows a phenomenon analogous to Bloch oscillations in a multi-junction Josephson circuit, the quantronium [8], which is a recently developed quantum bit circuit.

## 2  Bloch oscillations with the quantronium

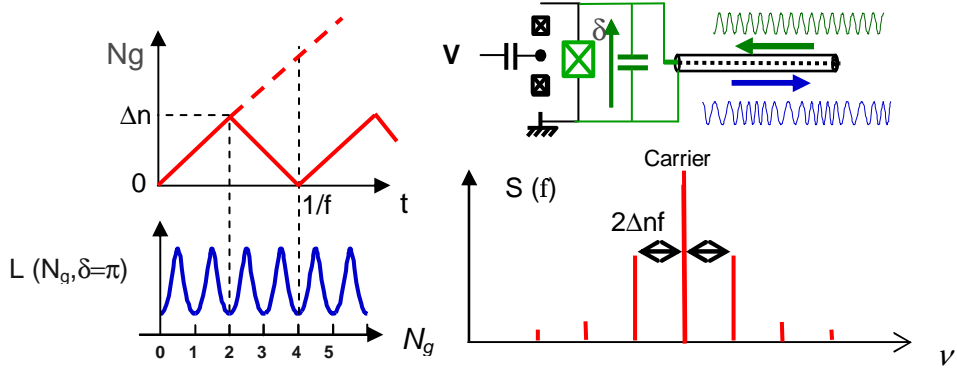

Fig (1). When a triangular gate charge modulation (top left) is applied to a quantronium (top right), its inductance varies periodically, which modulates the reflection coefficient of the device at microwave frequencies close to the plasma resonance of the larger junction on the right. When the extrema of the gate sweep correspond to integer or half-integer values of $N_g$, the modulation of the reflection factor is the same as in the case of a linear sweep of the gate charge (continuing dashed line). The spectrum of the reflected signal then presents sidebands shifted from the carrier by multiples of the Bloch frequency $2\Delta nf$. For an arbitrary periodic sweep of the gate charge, other sidebands are predicted at all frequency shifts $kf$.

The quantronium circuit, schematically shown in Fig. (1), can be figured as a Bloch transistor (i.e. two small junctions in series) in parallel with a larger junction. The circuit is connected to a microwave transmission line used for probing its impedance. The transistor Hamiltonian is controlled by the reduced charge on the gate electrode $N_g = C_g V/(2e)$, with $C_g$ the gate capacitance, and by the reduced flux $\delta = \phi/\varphi_0$, with $\phi$ the flux threading the loop, and $\varphi_0 = \hbar/2e$:

$$\hat{H} = E_C(\hat{N} - N_g)^2 - E_J \cos(\delta/2)\cos\hat{\theta} \qquad (1.2)$$

Here, the conjugate variables are the island pair number $N$ and phase $\theta$. All the circuit properties vary periodically with $N_g$ (period 1) and with $\delta$ (period $2\pi$). We consider here the case of a quantronium in which the large junction is loaded by an on-chip capacitor in order to place its plasma resonance frequency in the convenient $1-2\,\text{GHz}$ frequency range. Assuming the circuit stays in the ground state of Hamiltonian (1.2) with energy $\varepsilon(N_g,\delta)$, the effective inductance of the transistor $L = \varphi_0^2/(\partial^2\varepsilon/\partial\delta^2)$ varies with $N_g$ as shown in Fig (1), which modifies the reflection factor of the circuit. The calculations were performed for the parameters of the sample: $E_C \simeq 0.5\text{K}, E_J \simeq 1.0\text{K}$. If the gate charge $N_g$ could increase linearly in time, i.e. in the case of a constant gate current $I$, the circuit properties, such as the island potential or the impedance, would be modulated at the Bloch frequency $f_B = I/2e$. When a cw signal is injected on the microwave line, the spectrum of the reflected signal is expected to develop sidebands shifted from the carrier by the Bloch frequency. Harmonics are also expected since the modulation of the reflection coefficient is not purely sinusoidal. Of course it is impossible to maintain a constant current through the gate capacitor for a long time. It is nevertheless possible to carry out an experiment showing the Bloch oscillations predicted for a constant current, as now explained.

Let us consider the case of a triangular modulation with frequency $f$ and span $\Delta n$ of $N_g$ between two exactly integer or half-integer values, as shown in Fig. (1). In this situation, the modulation of the inductance, and thus of the reflection coefficient, is expected to be exactly the same as in the case of a constant gate current $i = 2\Delta n f (2e)$ because the inductance modulation is periodic, and symmetric around integer and half-integer values of $N_g$. The corresponding Bloch frequency is $f_B = 2\Delta n f$. With the inductance deduced from the ground-state energy [8], one easily calculates for an arbitrary periodic gate modulation the Fourier series giving the reflection factor

$$R(t) = \Sigma_k r_k \exp(2i\pi k f t). \tag{1.3}$$

The coefficients $r_k$ provide the reflected amplitudes at all frequencies $kf$ shifted from the carrier frequency. The Bloch sidebands shifted by $\pm f_B$ dominate when the triangular gate modulation is properly tuned, as predicted by the simple physical picture presented above. The spectrum $S(\nu)$ calculated following this procedure is in excellent agreement with a full calculation of the reflected signal in the linear excitation regime of the circuit.

## 3   Observation of Bloch oscillations

A quantronium sample, with an Al/AlOx/Al on-chip capacitor, was fabricated using electron-beam lithography, placed in a sample-holder fitted with microwave lines, and cooled down to 30 mK. The gate was connected to a 250 MHz bandwidth rf line, and the quantronium was connected both to a microwave injection line and to a measuring line through a circulator. The injected microwave power at the circuit level was chosen to maintain the phase dynamics in the harmonic oscillator regime. The reflected signal was sent through two decoupling circulators to a cryogenic amplifier at 4.2 K, and was finally amplified at room temperature. The effective gain of the measuring line was 76 dB. The signal was then either demodulated with the input cw signal in order to directly observe Bloch oscillations, which is possible at low Bloch frequencies, or sent to a spectrum analyzer. A series of spectra showing the Bloch lines is shown in Fig (2) for a properly tuned triangular gate modulation with $\Delta n$=10 and zero offset at different sweep frequencies. The Bloch lines have a narrow sub-Hz linewidth. We have checked the selection rules for the offset and the amplitude predicted from the triangular modulation patterns mimicking a linear evolution of the gate charge. When the offset is detuned, other non-Bloch sidebands appear, as calculated using Eq. (1.3). This experiment also revealed that the reflected signal randomly alternates between two values corresponding to gate charges shifted by one electron. This phenomenon is due to spurious quasiparticle states lying in the superconducting gap. This lack of robustness of the parity, commonly observed in Cooper pair devices, may be further enhanced when the gate charge is swept fast and over a wide range.

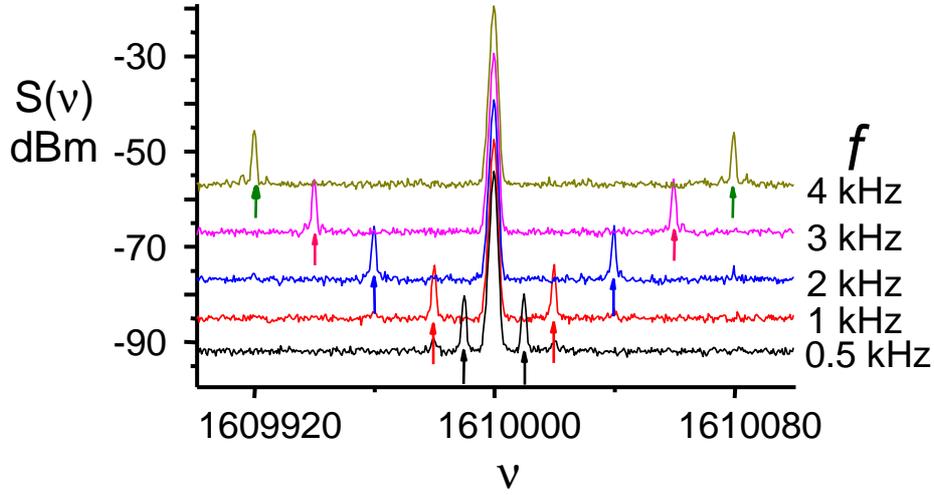

Fig (2). Spectrum of the reflected signal showing Bloch lines when a triangular gate charge modulation with zero offset and amplitude $\Delta n = 10$ is applied to the gate, for different frequencies $f$. Curves have been shifted vertically for clarity. The arrow indicates the predicted position of the Bloch lines at $f_C \pm 2\Delta n f$, with $f_C = 1.61\,\text{GHz}$ the carrier frequency. The second Bloch harmonic is visible on some traces.

The dependence of the sideband amplitude on $\Delta n$ is compared to predictions based on Eq. (1.3) on Fig. (3), for three frequencies of the triangular modulation. The agreement with the spectrum calculated with the inductance modulation deduced from the sample parameters is satisfactory at low frequency. The amplitude of an even sideband is maximum when it corresponds to a Bloch line. The dependence on $\Delta n$ was checked up to k=40. The agreement strongly degrades when the sweep frequency increases $f \geq 1\,\text{MHz}$. This discrepancy might arise from parity changes during the sweeps and from Zener transitions. High frequency operation of the quantronium should indeed be ultimately limited by Zener tunneling towards upper Bloch bands. The gap with the first excited band normally vanishes at the points $\{N_g = 1/2 \bmod(1), \delta = \pi\}$. In order to avoid this zero gap point, an asymmetry was introduced between the transistor junctions, yielding a designed gap frequency of 2 GHz. This allowed to operate the device close to $\delta = \pi$ where the gate-charge modulation of the inductance is maximum. The calculated Zener transition rate for this gap does not account however for the observed sweep frequency dependence of the spectrum in the present experiment.

## 4  Perspectives for metrology and for mesoscopic physics

The observation of Bloch oscillations in a Josephson circuit is a first step towards their use for the metrology of currents. A significant effort is presently devoted to close the quantum metrology triangle, which relates time, voltage and current units. The production of a current directly related to a frequency, or reversely, would allow to check consistency with the Josephson and Quantum Hall Effects. The currents presently produced by single electron pumps, in the pA range, are however too small compared to the currents needed in Quantum Hall Effect experiments, even when using cryogenic current comparators based on superconducting transformers. Metrological current sources in the 100 pA are clearly needed to improve the accuracy, presently limited at about $10^{-6}$.

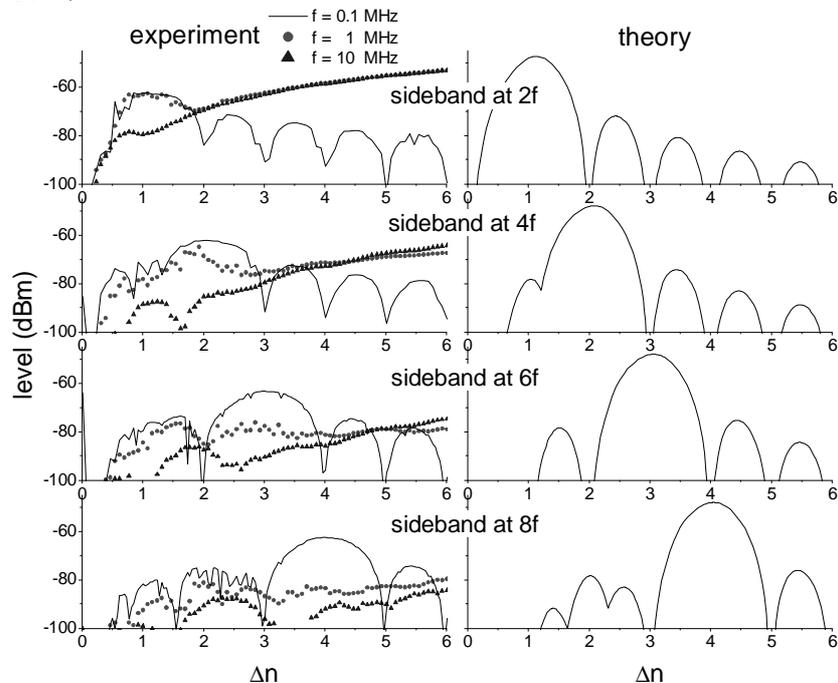

Fig (3). Comparaison of the measured sideband amplitude dependence (left) with the predicted one (right), for the even sidebands at three different sweep frequencies. A good agreement is found only at the lowest frequency, with a maximum when the sideband corresponds to a Bloch line. The amplitude dependence departs from the predicted one for sweep frequencies $\geq 1\,\mathrm{MHz}$.

The demonstration of Bloch oscillations at larger frequencies than achieved in this work, with the injection of a dc current, is thus an important goal. In order to inject such a current, the gate capacitor has to be replaced by an impedance large compared to the resistance quantum over a wide frequency range. Large chromium resistors, and linear arrays of junctions have already been used to achieve this goal, and the successful observation of a Bloch nose in the current-voltage characteristics of a small Josephson junction already proves that the Bloch oscillation regime is attainable [7]. The phenomenon of quantum phase slippage in an ultrasmall superconducting wire was also recently proposed to reach this regime [9].

Our experiment is also related to the electron counting experiment [10], which probes the voltage oscillations of a small electrode in a tunnel junction array with a radiofrequency single electron transistor (RFSET). In our experiment, the passage of a Cooper pair through the device induces a cyclic evolution of the microwave reflection coefficient. This time dependence results in Bloch lines in the spectrum, but its direct observation in the time domain is also possible, and was performed in the present experiment up to Bloch frequencies of a few kHz. Counting the number of periods completed in a given time, which could be achieved by processing the demodulation quadrature signals, would provide a direct measure of the current. The transposition of the injected current into the frequency domain may also be useful in mesoscopic physics since the shape of the Bloch lines is determined by the fluctuations of the injected current. The measured spectrum is thus related to statistical properties of the injected current, such as the third moment of its Full Counting Statistics.

In conclusion, we have observed Bloch oscillations in the quantronium circuit by microwave reflectometry when suitable signals are applied to the gate electrode of the device. The Bloch sidebands observed in the spectrum vary as predicted theoretically with the amplitude and offset of the gate modulation at small sweep frequency.

**Acknowledgments**

We acknowledge M. Devoret for discussions and for providing us with a sample for preliminary experiments, H. Mooij, D. Haviland, H. Grabert, and F. Hekking for discussions, and P.F. Orfila, P. Sénat, and M. Juignet for technical help. This work was supported by the european project Eurosqip.